\documentclass[twocolumn,showpacs,preprintnumbers,amsmath,amssymb]{revtex4}

\usepackage{graphicx}
\usepackage{dcolumn}
\usepackage{bm}

\begin{document}

\title{Quality factor of a matter-wave beam.}
\author{Fran\c{c}ois Impens}
\affiliation{SYRTE, Observatoire de Paris, CNRS, UPMC ; 61 avenue
de l'Observatoire, 75014 Paris, France.}
\date{\today}

\begin{abstract}
Imperfections in dilute atomic beams propagating in the paraxial regime and in potentials of cylindrical symmetry have been characterized experimentally through the measurement of a parameter analogous to a beam quality factor [Riou et al., Phys. Rev. Lett. 96, 070404 (2006)]. We propose a generalization of this parameter, which is suitable to describe dilute matter waves propagating beyond the paraxial regime and in fully general linear atom-optical systems. The presented quality factor shows that the atomic beam symmetry can be traded for a better transverse collimation.
\end{abstract}


\maketitle

\section{INTRODUCTION}

There is a strong analogy between the propagation of light and
matter waves~\cite{BordeHouches}. It is manifest in the paraxial
wave equation for the mode $U$ of an electromagnetic field
$\mathbf{E}(x,y,z)=U(x,y,z)e^{i(kz-\omega t)} \hat{\mathbf{e}}$
which takes the form of a bidimensional Schr\"odinger equation
\small
\begin{equation}
\label{eq:bidimensional light equation}
  2 i k \frac {\partial U (x,y,z)} {\partial z}  =- \left( \frac {\partial^2}
   {\partial x^2}+ \frac {\partial^2} {\partial y^2} \right) U(x,y,z)+ F(x,y) U(x,y,z) \,,
\end{equation}
\normalsize
with a function $F(x,y)$ depending on the local refraction index.
A spectacular achievement proving the similarity of the atomic
 and light
fields was the realization of a quasicontinuous matter beam
analog to a
laser~\cite{BordeMWA95a,BordeMWA95b,Ketterle97,Bloch99,
Gerbier01,Cennini03,Phillips99b,Guerin06}.
It has been shown recently that the extraction of a
quasicontinuous atom laser from a Bose-Einstein condensate  by a
weak RF~\cite{Riou06,Kohl05,Busch02,LeCoq01} or
Raman~\cite{Jeppesen07} outcoupling involves the propagation
through a discontinuous potential which degrades the beam
collimation. To quantify this effect, Riou \textit{et
al}.~\cite{Riou06} introduced a parameter analogous to a
quality factor. It allows one to estimate the the transverse width of
a cylindrical atomic beam falling in a uniform gravitational field
and in the paraxial regime, i.e., with atoms
strongly accelerated along one privileged direction.\\

However, an important difference between light and matter waves is
that the latter generally do not propagate in this paraxial
regime. This is typically the case for a condensate
expanding after a sudden trap shut down, which can be
viewed as a pulsed atom laser. Even for a weakly outcoupled
atom laser, the paraxial approximation is often valid only
after a certain propagation time~\cite{Riou06} during which the beam
must be treated in a more general framework. A comparison between the usual Schr\"odinger equation and
the paraxial wave equation (\ref{eq:bidimensional light equation})
shows that for these matter waves, time plays the role devoted to
the axis of propagation $Oz$ in paraxial optics, and that the
transverse space is three dimensional and spanned by
$\{\hat{\mathbf{e}}_x,\hat{\mathbf{e}}_y,\hat{\mathbf{e}}_z \}$
instead of $\{\hat{\mathbf{e}}_x,\hat{\mathbf{e}}_y \}$.\\

To extend the concept of matter beam quality beyond the paraxial
regime, it is necessary to reconsider its physical significance in
optics. For cylindrical light beams, a quality factor compares the
divergence of a non-ideal beam to a standard set by a perfect
Gaussian mode. It conveys the idea of degradation in the
collimation or, equivalently, that of diminution in phase-space
density. A relevant quality parameter should thus be left
invariant during the propagation in perfect linear optical systems
which preserve the beam collimation.\\

 Such systems are modelled
by a finite set of $ABCD$ matrices which describe linear
input-output relations in the light beam
phase-space~\cite{Bastiaans86}. Their atomic counterpart is the
quantum evolution under an Hamiltonian quadratic in position and
impulsion, which involves similar relations with $3 \times 3$
$ABCD$ matrices~\cite{BordeTheortool2001, BordeMetrologia2002}.
The phase space evolution of the atomic beam then also amounts to
a time-dependent map in the arguments of the Wigner
distribution~\cite{BordeHouches}
\begin{eqnarray}
W \left( \mathbf{r},\mathbf{p},t \right) = W \left( \:
\widetilde{D} (\mathbf{r} - \mathbf{\xi}) - \frac 1 m
\widetilde{B} (\mathbf{p} - m \mathbf{\phi}) \:  , \right. \nonumber \\
\left. - m \widetilde{C} (\mathbf{r} - \mathbf{\xi}) + \widetilde{A}
(\mathbf{p} - m \mathbf{\phi}) \: , \: t_0 \right) \,. \nonumber
\end{eqnarray}
The $\widetilde{}$ stands for the transposition, the vectors
$\mathbf{r},\mathbf{p}$ are the position and impulsion, the
matrices $A,B,C,D$ and vectors $\mathbf{\xi},\mathbf{\phi}$ are
time-dependent parameters determined in Ref.\cite{BordeTheortool2001}.
Thanks to the unimodularity of the $ABCD$ matrices, the
propagation under those quadratic Hamiltonians preserves the
phase-space density and as such do not alter the atomic beam
collimation. In this view, these Hamiltonians can be considered as
perfect aberrationless atom-optical systems and should thus leave
invariant  a  well-defined quality factor.\\

This is not the case for the parameter used so far to characterize
the quality of matter beams~\cite{Riou06,Jeppesen07}. This is
disturbing, since its current definition would sometimes lead to
ignoring possible improvements of the atom laser collimation through
non cylindrical atom-optical elements such as astigmatic atomic
lenses~\cite{Impens06b}. It seems thus useful to propose a new
quality factor for atom lasers which respects this invariance
requirement. This is the main purpose of this paper.

\section{QUALITY FACTOR OF A CYLINDRICAL BEAM}

Let us first reconsider the definition of a quality factor for a
cylindrical light beam, before extending this concept to atom
optics. Partially coherent light beams can be described by means
of a first-order field correlation function $\Gamma$ relating the
field amplitude at different points of planes transverse to the
direction of propagation $O_z$. For a cylindrical beam, only a
single transverse direction $O_x$ needs to be considered in the
correlation function
\begin{equation}
\Gamma(x_1,x_2,z)= \langle E(x_1,z) \: E^*(x_2,z) \rangle \,.
\nonumber
\end{equation}
The Wigner transform of this function provides a phase space
picture of the beam
\begin{equation}
W(x,k_x,z)= \int {\rm d}x' \: \Gamma(x+ \frac {x'} {2} ,x- \frac {x'}
{2},z) \: e^{ -i k_x x'} \,, \nonumber
\end{equation}
which can be used to define moments in the transverse position and
wave vectors
\begin{equation}
\langle m(x,k_x,z) \rangle = \int {\rm d} x \: {\rm d} k_x \: m(x,k_x,z)  \:
W(x,k_x,z) \,. \nonumber
\end{equation}
Because the beam is cylindrical and transverse $\langle x
\rangle=\langle k_x \rangle=0$. The moments $\Delta x=
\sqrt{\langle x^2 \rangle}$ and $\Delta k_x= \sqrt{\langle k_x^2
\rangle}$ are, respectively, the transverse
squared width and wave-vector dispersions of the beam.\\

 Optimal collimation is
achieved with a perfectly coherent Gaussian mode, for which the
position and direction moments
 verify at the waist
\begin{equation}
 \Delta x |_{w,\mbox{Gaussian}} \: \Delta
k_x |_{w,\mbox{Gaussian}}  = \frac {1} {2} \,. \nonumber
\end{equation}
Partially coherent cylindrical light beams have a lessen
collimation quantitatively estimated thanks to a quality factor
$M^2$~\cite{Siegman91}. This parameter can be expressed as a
combination of moments left invariant during the propagation in
aberrationless  cylindrical optical  systems
\begin{equation}
\label{eq:unidimensional moment invariant} \frac {M^2} {2} \: = \:
\sqrt{\langle x^2 \rangle|_{z} \:  \langle k_x^2 \rangle|_{z} \: -
\: \langle x \:k_x \rangle|_{z}^2} \,.
\end{equation}
At the beam waist, this expression reduces to the product of
transverse squared width and divergence which has a clear
interpretation in terms of phase-space dispersion
\begin{equation}
\label{eq:definition optical quality1} \frac {M^2} {2} =  \Delta x
|_{w} \: \Delta k_x |_{w} \,.
\end{equation}
This is indeed the definition which has been adopted to express
the quality of a matter beam~\cite{Riou06,Jeppesen07}, once taken
into account De Broglie relation between momentum and wave vector
$\mathbf{p}=\hbar \mathbf{k}$
\begin{equation}
\label{eq:definition M1}
 M_{1}^2 = \frac  {2} {\hbar} \Delta x |_{w} \: \Delta p_x
|_{w} \,.
\end{equation}
This definition is consistent if one considers only matter-waves
and atom-optical systems with cylindrical symmetry, and a
propagation in the paraxial regime: the parameter $M_{1}^2$ is
then unaltered by the propagation in aberrationless systems.\\

However, these conditions are far too restrictive to characterize the
matter beams in most experiments, which generally involve
potentials without any symmetry and a matter-wave propagation
outside the paraxial regime. A three-dimensional quality factor
$M_3$, invariant under the whole set of possible $ABCD$ matrices,
is then required. Following previous treatments of optical and
quantum mechanical invariants~\cite{Serna91,Dragt92,Simon00}, we
now examine how to properly define this parameter.

\section{QUADRATIC PROPAGATION INVARIANT FOR MATTER BEAMS}

Our goal is to replace the  parameter $M_1$ by an $ABCD$-invariant
combination of moments which still provides an insight on the matter
 beam divergence. In the unidimensional case, the price to pay to express the
product $\Delta x |_{w} \: \Delta k_x |_{w}$  (\ref{eq:definition
optical quality1}) as an invariant expression
(\ref{eq:unidimensional moment invariant}) was to introduce an
additional moment $\langle x \:k_x \rangle$ mixed in wave vector
and position. This term can be viewed as an element of a variance
matrix. Such matrices have been introduced in
optics~\cite{Serna91}, and they can also characterize a partially
coherent atomic beam~\cite{AntoineThese}:
\begin{equation}
V = \left( \begin{array} {cc} \Delta_{\mathbf{r} \mathbf{r}} &
\Delta_{\mathbf{r} \mathbf{v}} \\
\widetilde{\Delta}_{\mathbf{r} \mathbf{v}} & \Delta_{\mathbf{v}
\mathbf{v}}
\end{array} \right) \,. \nonumber
\end{equation}
The matrices $\Delta_{\mathbf{a} \mathbf{b}}$ are
\begin{equation}
\Delta_{\mathbf{a} \mathbf{b}} = \left( \begin{array} {ccc}
\langle a_x b_x \rangle &
\langle a_x b_y \rangle  & \langle a_x b_z \rangle \\
\langle a_y b_x \rangle &
\langle a_y b_y \rangle  & \langle a_y b_z \rangle \\
\langle a_ z b_x \rangle & \langle a_z b_y \rangle  & \langle a_z
b_z \rangle
\end{array} \right)  \nonumber
\end{equation}
with the vectors $\mathbf{a},\mathbf{b}=\mathbf{r},\mathbf{p}/m$.
We now derive the variance evolution when the atomic field
propagates under an Hamiltonian which is quadratic in position and
impulsion
\begin{equation}
\label{eq:introduction hamiltonien quadratique general}
H(\hat{\mathbf{r}},\hat{\mathbf{p}}) = \frac{
\hat{\widetilde{\mathbf{p}}} \beta(t) \hat{\mathbf{p}}} {2 m} -
\hat{\widetilde{\mathbf{r}}} \alpha(t) \hat{\mathbf{p}} - \frac
{m} {2} \hat{\widetilde{\mathbf{r}}} \gamma(t) \hat{\mathbf{r}} -
m \mathbf{g}(t) \cdot \hat{\mathbf{r}}
 +\mathbf{f}(t) \cdot \hat{\mathbf{p}} \,.
\end{equation}\\
$\alpha(t)$, $\beta(t)$ and $\gamma(t)$ are $3 \times 3$ matrices;
$\mathbf{f}(t)$ and $\mathbf{g}(t)$ are three dimensional vectors.
It follows from the equations of motion for the position and
impulsion operators in the Heisenberg picture that the variance
matrix $V$ satisfies
\begin{equation}
\label{eq:variancematrix evolution}
\frac {d V} {dt} = \Gamma V + V \Gamma \quad \quad \Gamma(t) =
\left( \begin{array} {cc} \alpha(t) & \beta(t) \\
\gamma(t) & \alpha(t) \end{array} \right) \,. \nonumber
\end{equation}
The integration of this last equation involves the time-ordering
operator $T$ and the atom-optical $ABCD$
 matrix~\cite{BordeGRG2004}:
\begin{eqnarray}
 M(t,t_0)& = & \left(
\begin{array}{cc}
A(t,t_0) & B(t,t_0)  \\
C(t,t_0) & D(t,t_0)  \\
\end{array} \right) \nonumber \\
&  = &  T \: \:  \exp \left[ - \int_{t_0}^{t}
{\rm d} t' \left(
\begin{array}{cc}
\alpha(t') & \beta(t')  \\
\gamma(t') & \alpha(t')  \\
\end{array} \right) \right]  \,. \nonumber
\end{eqnarray}
The evolution of the variance matrix in the quadratic
Hamiltonian~(\ref{eq:introduction hamiltonien quadratique
general}) is then similar to the transformation laws in optics
\begin{equation}
\label{eq:variance propagation}
 V(t)\: \: = \: \: M(t,t_0) \:  V(t_0) \:
\widetilde{M}(t,t_0) \,.
\end{equation}
 The most straightforward
attempt to generalize the unidimensional quality factor $M_1^2$ to
 three dimensions would consist in considering norms and scalar
products in Eq.(\ref{eq:unidimensional moment invariant}) instead of
single coordinates
\begin{equation}
 \sqrt{\langle x^2+y^2+z^2 \rangle
\langle p_x^2+p_y^2+p_z^2 \rangle - \langle \mathbf{r} \cdot
\mathbf{p} \rangle^2} \,. \nonumber
\end{equation}
At an instant satisfying $\langle \mathbf{r} \cdot \mathbf{p}
\rangle =0$, the multidimensional equivalent of a ``waist'', this
expression would express the greater phase space occupied by the
matter beam. Unfortunately, this quantity is not left invariant in
the propagation under a general quadratic Hamiltonian.\\

Nonetheless, it is possible to define a quality factor which
respects the invariance requirement and still provides a
meaningful insight into the beam phase space. We use the fact that
the $ABCD$ matrices associated with matter wave propagation are
symplectic, i.e., they verify at all times the following
relation
\begin{equation}
\label{eq:definition symplectic} M(t,t_0)^{-1} = L \: M(t,t_0) \:
\widetilde{L} \Longleftrightarrow L = \: \widetilde{M}(t,t_0) \: L
\: M(t,t_0) \,,
\end{equation}
with the matrix
\begin{equation}
 L = i \left(
\begin{array} {cc} 0 & -1 \\ 1 & 0 \end{array}  \right) \,. \nonumber
\end{equation}
 This symplectic structure  can indeed be used to
generate a family of invariants \cite{Serna91,Simon00,Dragt92}.
The lowest order of this invariant family indeed generalizes the
definition of the current matter beam quality factor
\begin{equation}
\label{eq:definition invariant} M_3^4(t) =  \frac {4 m^2} {3
\hbar^2} \mbox{Tr} \left[ V(t) \:L \: V(t) \: L \right] \,.
\end{equation}
The constant is adjusted to yield $M_3=1$ for a Gaussian matter
wave. Proving the invariance of this parameter is simple and
identical to optics~\cite{Serna91}. Combining relation
(\ref{eq:variance propagation}) describing the variance matrix
propagation with the cyclic property of the trace, the
definition~(\ref{eq:definition invariant}) can be recast as
\small
\begin{equation}
M_3^4(t)  =  \frac {4 m^2} {3 \hbar^2} \mbox{Tr} \left[ V(t_0)
\widetilde{M}(t,t_0) L \widetilde{M}(t,t_0)   V(t_0)
\widetilde{M}(t,t_0)  L  M(t,t_0) \right] \,. \nonumber
\end{equation}
\normalsize
The symplectic relation (\ref{eq:definition symplectic}) then
gives the desired invariance of $M_3(t)$:
\begin{equation}
M_3^4(t)  =  \frac {4 m^2} {3 \hbar^2} \mbox{Tr} \left[ V(t_0) \:
L \:  V(t_0) \: L \right]= M_3^4(t_0) \,.
\end{equation}
Its expression in terms of position and impulsion momenta is
\small
\begin{eqnarray}
\label{eq:definition invariant2}
 M_3^4  & = & \frac {4} {3 \hbar^2} \left[
  \langle x^2 \rangle \langle p_x^2 \rangle -
\langle x \: p_x \rangle^2+ \langle y^2 \rangle \langle p_y^2
\rangle - \langle y \: p_y \rangle^2 \right.  \nonumber \\
& + & \langle z^2 \rangle \langle
p_z^2 \rangle
   -   \langle z \: p_z \rangle^2
 +2  (\langle x y \rangle \langle p_x p_y \rangle
 - \langle x p_y \rangle \langle y p_x
 \rangle) \nonumber \\
 & + & \left. 2 (\langle x z \rangle \langle p_x p_z \rangle
 - \langle x p_z \rangle \langle z p_x \rangle)
 +2 (\langle y z \rangle \langle p_y p_z \rangle- \langle
y p_z \rangle \langle z p_y \rangle) \right] \frac {} {} \,. \nonumber \\
\end{eqnarray}
\normalsize
This expression can be rewritten in a more compact form as a sum
of moments along single and multiple directions:
\begin{eqnarray}
 M_3^4 & = & \frac {1} {3} \left[ Q_x+Q_y+Q_z-A_{xy}-A_{xz}-A_{yz}
 \right]
  \nonumber \\
Q_{\eta} & = &  \frac {4} { \hbar^2} \langle \eta^2 \rangle \langle
p_{\eta}^2 \rangle  \nonumber \\
 A_{\eta \: \eta'} & = &
\frac {8} { \hbar^2} ( \langle \eta \: p_{\eta} \rangle \langle
\eta'
 p_{\eta'} \rangle -\langle \eta \: \eta' \rangle
\langle p_{\eta} p_{\eta'} \rangle )
\end{eqnarray}
The quantities $Q_{\eta}$ correspond to the fourth power of the
previous matter-wave quality factor $M_1$ considered along the
three spatial directions, while the terms $A_{\eta \: \eta'}$
reflect
 correlations between different directions. If one chooses the
 coordinate system along the beam principal axis, the position
  momenta satisfy $\langle \eta \: \eta'
 \rangle=0$ for  $\eta \neq \eta'$, yielding a simpler expression
 for $A_{\eta \: \eta'}$:
\begin{equation}
A_{\eta \: \eta'} =   \frac {8} { \hbar^2} \langle \eta \:
p_{\eta'} \rangle \langle \eta'
 p_{\eta} \rangle \nonumber
\end{equation}
  These two families of parameters
  have different physical significance and obey different constraints. The
  parameters $Q_{\eta}$ are a manifestation of the beam divergence
and are bounded below by the Heisenberg principle: $Q_{\eta} \geq
1$. The terms $A_{\eta \: \eta'}$  reflect the beam asymmetry and
can be of either sign, they cancel for spherical clouds. As
previously announced,
 a Gaussian matter wave satisfies $M_3=1$.\\

This atomic beam quality factor thus reflects the departure
from a fundamental Gaussian mode: a beam with a quality
factor $M_3 \gg 1$ needs to be expanded onto several modes,
which is likely to degrade the fringe pattern in an atomic
interference experiment or to add additional noise in the
atomic beam~\cite{Johnsson07}. In principle, if these modes were put into
a fully coherent and accurately controlled superposition, a
high value of the quality factor would merely induce an additional
complexity in the fringe pattern. This ideal situation
is, however, not encountered in practice, since the coefficients
of the mode decomposition are generally not accessible.
For most precision interferometric measurement, a
simple $\mbox{TEM}_{00}$ mode is indeed highly preferable. In the LIGO
experiment, the beam quality factor of the laser involved is
maintained at a value $M\simeq1$ thanks to a mode-cleaning step
which filters out high order modes~\cite{Abramovici92,Uehara97}. This quality factor
requirement, reflecting the control on the beam structure,
should also apply to accurate interferometers involving atom
lasers.

It seems in general more advantageous to minimize the quantities
$Q_{\eta}$ which are directly related to the beam collimation. In
this respect, the fact that the invariant quantity be $M_3$
instead of the coefficients $Q_{\eta}$ has a practical interest.
Considering for instance a cylindrical beam with given
$Q_x=Q_y=Q_r=  M_1^4$ and $Q_z$, it is possible to reduce $Q_r$
while keeping the quality factor $M_3$ constant by increasing
$Q_z$ or by generating negative asymmetrical parameters $A_{\eta
\: \eta'}$~\cite{Serna92}. One can thus trade some longitudinal
collimation or cylindrical symmetry for a better transverse
collimation. The corresponding $ABCD$ transformation could be
implemented by astigmatic atom-optical lenses such as
electromagnetic waves with an asymmetric
wave-front~\cite{Impens06b}. This possibility of improvement is
ignored in Ref.~\cite{Riou06,Jeppesen07}, which assume that the transverse divergence measured at the waist
(i.e. $M_1^2$) is an upper bound for the atom laser
collimation in the subsequent propagation.

\section{CONCLUSION}

A quality factor has been defined for matter waves [$M_3$, Eq.(\ref{eq:definition invariant2})],
which addresses the general
propagation of an atom laser. It generalizes the currently adopted
beam quality factor [$M_1$, Eq. (\ref{eq:definition M1})], indeed
only appropriate to describe the propagation of matter-waves in
the paraxial regime and in cylindrical potentials, and which can
lead to underestimate the optimal collimation of an atom laser.
Higher order invariants proportional to $I_k = \left( \frac {m}
{\hbar} \right)^{2k} \: \mbox{Tr} \left[ V(t) \:L \: V(t) \: L
\right]^k$ could also be considered to describe the atomic beam, but
they do not provide
the same insight into the  phase-space density.\\

\textit{Note:} Since the publication of this paper, a different nonlinear matter wave quality
factor has been proposed~\cite{Impens09c}, which is preserved in presence of mean-field interactions. However, this nonlinear parameter is suitable only for uniform and paraxial atomic beams propagating in transverse potentials of cylindrical symmetry. The generalizations of the matter wave quality factor presented in this paper and in Ref.~\cite{Impens09c} are indeed complementary: they enable one to apply the concept of beam quality beyond the paraxial and beyond the linear regime of propagation respectively. The non-paraxial and the non-linear matter wave quality factors can thus be fruitfully applied in different contexts.\\

\section*{ACKNOWLEDGEMENTS}
The author is greatly indebted to Christian J. Bord\'{e} for
enlightening discussions on atom optics. He thanks Emeric De Clercq for manuscript reading. This work was supported by DGA.

\end{document}